\documentclass[aps,prl,twocolumn,superscriptaddress,amsfonts,amssymb,amsmath]{revtex4-1}
\usepackage{txfonts}
\usepackage{multirow}
\usepackage{color}
\usepackage{dcolumn}
\usepackage{ulem}
\usepackage{graphicx}

\newcolumntype{d}[1]{D{.}{\cdot}{#1}}
\newcolumntype{.}{D{.}{.}{-1}}
\newcolumntype{,}{D{,}{,}{-1}}

\begin{document}

\title{Hindered Quadrupole Order in PrMgNi$_{4}$ with a Nonmagnetic Doublet Ground State}

\author{Yuka Kusanose} \affiliation{Graduate School of Advanced Sciences of Matter, Hiroshima University, Higashi-Hiroshima 739-8530, Japan}
\author{Takahiro Onimaru} \affiliation{Graduate School of Advanced Sciences of Matter, Hiroshima University, Higashi-Hiroshima 739-8530, Japan}
\author{Gyeong-Bae Park} \affiliation{Graduate School of Advanced Sciences of Matter, Hiroshima University, Higashi-Hiroshima 739-8530, Japan}
\author{Yu Yamane} \affiliation{Graduate School of Advanced Sciences of Matter, Hiroshima University, Higashi-Hiroshima 739-8530, Japan}
\author{Kazunori Umeo} \affiliation{Cryogenics and Instrumental Analysis Division, N-BARD, Hiroshima University, Higashi-Hiroshima, Hiroshima 739-8526, Japan}
\author{Toshiro Takabatake} \affiliation{Graduate School of Advanced Sciences of Matter, Hiroshima University, Higashi-Hiroshima 739-8530, Japan}
\author{Naomi Kawata} \affiliation{Natural Science Center for Basic Research and Development, Hiroshima University, Higashi-Hiroshima, Hiroshima 739-8526, Japan}
\author{Tsutomu Mizuta} \affiliation{Graduate School of Science, Hiroshima University, Higashi-Hiroshima, Hiroshima 739-8526, Japan}

\begin{abstract}
Structural, transport and magnetic properties of single-crystalline samples of a praseodymium-based cubic compound PrMgNi$_4$ were studied. 
The single-crystal X-ray structural analysis revealed that Mg atoms are substituted for the Pr atoms at the 4$a$ site by 4.5\%. 
The $\chi(T)$ data follow the Curie-Weiss law with an effective moment for the Pr$^{3+}$ ion.
The magnetic specific heat divided by temperature, $C_{\rm m}/T$, shows a broad maximum at around 3 K, which is reproduced by a two-level model with a ground state doublet.
On cooling below 1 K, $C_{\rm m}/T$ approaches a constant value, which behavior is reproduced by a random two-level model. 
The twofold degeneracy of the ground state is lifted by symmetry lowering due to the substituted Mg atoms for the Pr atoms or strong hybridizations between the 4$f^2$ electron states and conduction bands, which hinders the long-range quadrupole order.
\end{abstract}

\maketitle


Praseodymium-based cubic compounds with a 4$f^2$ configuration have attracted much attention because a variety of phenomena manifest themselves at low temperatures. 
Pr-filled skutterudites exhibit correlated electronic phenomena due to strong hybridization between the 4$f$ electron and conduction electrons ($c$-$f$ hybridization), e.g., heavy-fermion superconductivity, a metal-insulator transition, and higher-order multipolar ordering.\cite{Bauer02,Sekine97,Kiss06,Sato09} 
These phenomena are attributed to the interplay between the conduction electrons and multipolar degrees of freedom in quasi-degenerated ground states of the Pr ions under the crystalline electric field (CEF).

In case the CEF ground state is the non-Kramers doublet of the Pr ion under cubic symmetry, the magnetic dipole is quenched but electric quadrupole and magnetic octupole become active.
Thereby, the multipoles in the non-Kramers doublet could govern the low-temperature properties.
A typical system is PrPb$_3$, which exhibits an antiferroquadrupole (AFQ) order at $T_{\rm Q}$ $=$ 0.4 K\cite{Bucher72,Morin82,Tayama01}. Below $T_{\rm Q}$, the quadrupoles are modulated sinusoidally by the correlation between the quadrupoles and the conduction electrons\cite{Onimaru05,Yamamura19}.
Another example is PrInNi$_4$, which presents a ferromagnetic transition at $T_{\rm C}$ $=$ 0.75 K.\cite{Walker06,Tsujii02,Tsujii03}
The nonmagnetic ground state doublet could mix with the excited magnetic triplet by the ferromagnetic exchange interaction inducing a magnetic moment in the ground state which orders ferromagnetically.

On the contrary, PrAg$_2$In and PrMg$_3$ exhibit no quadrupole order.\cite{Yatskar96,Tanida06,Morie09}
The absence of the quadrupole order was discussed by considering the Kondo effect due to strong $c$-$f$ hybridization, because the specific heat divided by temperature, $C/T$, is largely enhanced at low temperatures.
However, the thermoelectric power is not enhanced and the absolute value is comparable to that of the La counterpart, indicating the conduction electrons at the Fermi level could not correlate strongly with the 4$f^2$ electrons.\cite{Isikawa07,Isikawa09}
Moreover, in the specific heat measurements using a high-quality single crystal of PrAg$_2$In with the residual resistivity ratio of RRR $=$ 14, a cusp-type anomaly manifests itself at 0.33 K, indicating the relationship between the sample quality and occurrence of a phase transition.\cite{Sato14}
Thereby, the absence of the quadrupole order may result from symmetry lowering due to the atomic disorder inherent in the Heusler-type structure.\cite{Isikawa07,Isikawa09}

Recently, coexistence of quadrupole orders and superconductivity has been observed in non-Kramers doublet systems Pr$T$$_2$Zn$_{20}$ ($T$ $=$ Ir, Rh)\cite{Pr1-2-20,Onimaru10,Onimaru11,Onimaru12}, and Pr$T$$_2$Al$_{20}$ ($T$ $=$ Ti, V),\cite{Sakai11,Sakai12,Matsubayashi12,Tsujimoto14} which suggests the superconducting pair could be mediated by quadrupole fluctuations. 
Furthermore, non-Fermi liquid (NFL) behaviors of the electrical resistivity $\rho(T)$ and the specific heat $C(T)$ were observed not only in PrIr$_2$Zn$_{20}$ and PrRh$_2$Zn$_{20}$\cite{Onimaru16,Yoshida17} but also in a diluted Pr system Y(Pr)Ir$_2$Zn$_{20}$\cite{Yamane18a}, indicating possible manifestation of the quadrupole Kondo effect.\cite{Tsuruta15,Cox98}

The intensive works on the 4$f^2$ systems described above have revealed a rich variety of phenomena due to the interactions between the multipolar degrees of freedom of the 4$f^2$ electrons and conduction electrons.
Deeper understanding of the multipole-driven phenomena requires a search for a new Pr-based compound with the non-Kramers doublet ground state.
In the present work, we focused on PrMgNi$_4$ crystallizing in the cubic MgSnCu$_4$-type structure\cite{Kadir02}.
We analyzed the crystal structure and studied transport and magnetic properties at low temperatures down to 0.1 K.
The results are discussed to understand how the active quadrupoles in the doublet could be involved in the formation of the unusual ground state.


\begin{table*}[t]
\caption{Crystallographic parameters for PrMgNi$_4$ determined at 173 K. $U_{\rm eq}$ is the isotropic displacement parameter defined as 1/3 of the trace of the orthogonalized $U_{ij}$ tensor.}
\label{tbl2}
\begin{center}
\footnotesize
\begin{tabular}{lllllll}
\hline
\multicolumn{5}{l} {Cubic MgSnCu$_4$-type} \\
\multicolumn{5}{l} {Space group: $F\bar{4}3m$ (No. 216)} \\
\multicolumn{5}{l} {$a$ $=$ 7.0866(11) \AA, $V$ $=$ 355.89(17) \AA$^3$, $Z$ $=$ 4} \\
\hline
Atom & Site & $x$ & $y$ & $z$ & Occupancy & $U_{\rm eq}$ (\AA$^2$) \\ 
\hline
Pr & 4$a$ & 0 & 0 & 0 & 0.955(7) & 0.0038(2) \\ 
Mg & 4$a$ & 0 &  0 &  0 & 0.045(7) & 0.0038(2) \\ 
Mg & 4$c$  &  0.25 & 0.25  & 0.25 & 1 & 0.0073(7)  \\ 
Ni & 16$e$ & 0.37643(4) & 0.37643(4) & 0.62357(4)  & 1 & 0.0040(3) \\ 
\hline
\end{tabular}
\end{center}
\end{table*}



We have grown single crystals of PrMgNi$_4$ and LaMgNi$_4$ by the Mg self-flux method. 
The binary alloys PrNi$_4$ (LaNi$_4$) were arc-melted into ingots.
The ingot and Mg shot were doubly sealed in a molybdenum crucible and a quartz ampoule.
The ampoule was heated up to 1100 $^{\circ}$C in an electric furnace and then cooled down slowly.
At 700 {$^{\circ}$C, the ampoule was quickly removed from the furnace and centrifuged to remove the molten Mg flux.
Typical size of the single crystals is about 3 mm in length.

The samples were characterized by the electron-probe microanalysis (EPMA) and powder X-ray diffraction technique.
The atomic compositions determined by averaging over 10 different regions for each crystal with a JEOL JXA-8200 analyzer.
Assuming full occupancy of the Ni site, we obtained the compositions as Pr$_{0.97(9)}$Mg$_{1.09(5)}$Ni$_{4}$ and La$_{0.97(4)}$Mg$_{1.14(1)}$Ni$_{4}$, where excess Mg atoms were observed with respect to the stoichiometric composition of 1:1:4. 
No impurity phase was detected in view of the backscattered electron images and the powder X-ray diffraction patterns. 


The single-crystal X-ray structural analysis was performed at 173 K with Mo $K{\alpha}$ radiation, $\lambda$ $=$ 0.071073 nm, monochromated by a multilayered confocal mirror using a Bruker APEX-II ULTRA CCD area-detector diffractometer at N-BARD, Hiroshima University. The details of the measurement, data collection, and refinement are described in Supplemental Materials\cite{supplemental}. 

To identify the atomic coordination correctly, we evaluated a reliable factor defined as $R_{wp}$ $=$ $[\sum_{i}w_{i}(y_{i} - f_{i})^2/\sum_{i}w_{i}y_{i}^2]^{1/2}$, where $y_i$  and $f_{i}$ are the observed and calculated intensity of the X-ray diffraction, respectively, and $w_{i}$ is statistical weight for a Bragg peak labeled $i$.
First, we assumed the ordered structure, where the Pr, Mg, and Ni atoms exclusively occupy the 4$a$, 4$c$, and 16$e$ sites, respectively\cite{Kadir02}.
Thereby, the value of $R_{wp}$ $=$ 0.0213 was obtained.
Next, we considered possible exchange between Pr and Mg atoms, because the exchange of Ce and Mg in the isostructural CeMgNi$_4$ was proposed by the structural analyses of powder X-ray and neutron diffraction patterns\cite{Roquefere09}. A smaller value of $R_{wp}$ $=$ 0.0166 was obtained, where 3.6\% of Pr(Mg) at the 4$a$(4$c$) site are exchanged by Mg(Pr).
Lastly, following a report that excess $R$ atoms are substituted for Mg in $R$MgNi$_4$ ($R$ $=$ Ce, Gd, Er, and Lu)\cite{Linsinger11}, we adopted a model where excess Pr(Mg) atoms are substituted for Mg(Pr) atoms at the 4$c$(4$a$) site. 
Here, we assumed that the substituted site is fully occupied by Pr and Mg with equivalent displacement parameters.
The reliable factor became smallest $R_{wp}$ $=$ 0.0144, when excess Mg atoms by 4.5(7)\% are substituted for Pr at the 4$a$ site.
In $R$MgNi$_4$ ($R$ $=$ Ce, Gd, Er, and Lu), it was reported that excess $R$ atoms by 5-41\% are substituted for Mg at the 4$c$ site \cite{Linsinger11}.
This opposite occupations may be related to the lattice parameters, which are much smaller than that for PrMgNi$_4$.
However, it is an open question how the lattice parameter is related to the atomic substitution.
The crystallographic parameters of PrMgNi$_4$ obtained by the model with the excess Mg atoms are summarized in Tables \ref{tbl2}.

\begin{figure}
\centering
\includegraphics[width=10pc]{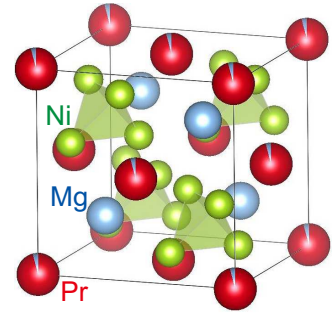}
\caption{(Color online) Cubic MgSnCu$_4$-type crystal structure of PrMgNi$_4$ with the space group of $F\bar{4}3m$\cite{Kadir02}. Large (red) spheres denote Pr atoms, the small (blue) ones Mg atoms, and the tetrahedra are formed by Ni atoms shown with the (green) spheres. The Pr sites are partially substituted by Mg atoms. 
}
\label{f1}
\end{figure}

As shown in Fig. \ref{f1}, the Pr atoms occupy the face-center position of the cubic unit cell. 
The point group of the Pr site is the cubic $T_d$.
As described above, the analysis revealed that the excess Mg atoms occupy the Pr sites by 4.5\%, which is moderately consistent with the atomic composition Pr$_{0.97(9)}$Mg$_{1.09(5)}$Ni$_{4}$ measured by the EPMA.
The substitution of Mg for Pr is allowed probably because the length between the Pr site and the nearest Ni atom, $d_{\rm Pr-Ni}$ $=$ 2.9410(5) \AA, is close to that of the Mg site, $d_{\rm Mg-Ni}$ $=$ 2.9350(5) \AA, although the ionic radius of the Mg$^{2+}$ ion is smaller than that of the Pr$^{3+}$ ion.
The smaller ionic radius of the Mg ion could give rise to the larger $U_{\rm eq}$ of Mg at the 4$c$ site than those of the other sites as shown in Table \ref{tbl2}.
The partial substitution of Mg for Pr must lower the local symmetry of the cubic Pr site, which could lift the twofold degeneracy of the ground state doublet of the Pr ion as will be discussed later.


The electrical resistance was measured by a standard four-probe AC method in the temperature ranges 3--300 K and 0.1--4 K, respectively, with a laboratory-built system installed in a Gifford-McMahon-type refrigerator and a commercial Cambridge Magnetic Refrigerator mFridge. 
The magnetization was measured from 1.8 to 300 K in magnetic fields for $B$ $\le$ 5 T by using a commercial superconducting quantum interference device (SQUID) magnetometer (Quantum Design, MPMS). 
The specific heat was measured by the thermal relaxation method using a Quantum Design physical property measurement system for 0.4 $<$ $T$ $<$ 20 K and the mFridge for 0.1 $<$ $T$ $<$ 0.6 K.


Temperature variations of the electrical resistivity $\rho(T)$ of $R$MgNi$_4$ ($R$ $=$ Pr and La) are shown in Fig. \ref{f2}.
The electric current was applied along the [100] axis of the single-crystalline samples.
The values of the residual resistivity ratio (RRR) evaluated by $\rho$(300 K)$/$$\rho$(2.8 K) are 3.0 and 2.8 for $R$ $=$ Pr and La, respectively.
On cooling from 300 K, $\rho(T)$ for $R$ $=$ La monotonically decreases and approaches a constant value of $\rho_0$ $=$ 15 ${\mu}{\Omega}$ cm at 2.8 K.
On the other hand, $\rho(T)$ for $R$ $=$ Pr exhibits a shoulder at around 10 K as shown with the arrow.
This anomaly may arise from the scattering of the conduction electrons by the thermal excitation of the CEF levels of the Pr ion\cite{Abou75}.

\begin{figure}
\centering
\includegraphics[width=17pc]{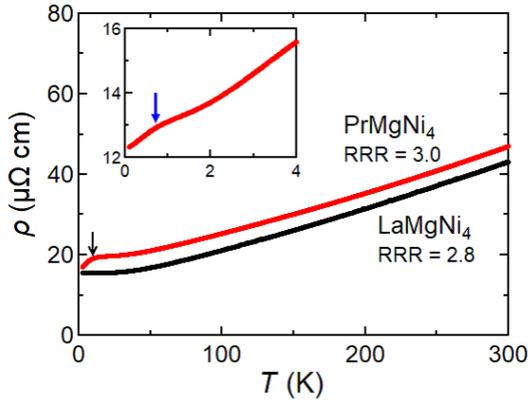}
\caption{(Color online) Temperature dependence of the electrical resistivity $\rho(T)$ of PrMgNi$_4$ and the nonmagnetic counterpart LaMgNi$_4$. The values of the residual resistivity ratio RRR are 3.0 for PrMgNi$_4$ and 2.8 for LaMgNi$_4$. The arrow indicates a shoulder at around 10 K. The inset shows the expanded $\rho(T)$ data of PrMgNi$_4$ for $T$ $\le$ 4 K, where the (blue) arrow indicates a broad shoulder at 0.8 K.}
\label{f2}
\end{figure}

Figure \ref{f3} shows the temperature dependence of the magnetic susceptibility $\chi$ of PrMgNi$_4$ measured in the magnetic field of $B$ $=$ 1 T applied along the [100] direction. 
The value of $\chi$ was calculated  by using the atomic composition determined by the EPMA. The inset shows $\chi^{-1}(T)$, which follows a modified Curie-Weiss equation $\chi(T)$ $=$ $C$$/$($T$ $-$ $\theta_{p}$) $+$ $\chi_{0}$, where $C$, $\theta_{p}$, and $\chi_{0}$ are the Curie constant, paramagnetic Curie temperature, and temperature independent susceptibility, respectively.
As shown with the solid curve, $\chi(T)$ between 20 and 300 K can be fitted by the above equation  with $\theta_{\rm p}$ $=$ $-$6.5 K and $\chi_{0}$ $=$ 6.8 $\times$10$^{-4}$ emu/mol.
The negative value of $\theta_{\rm p}$ indicates the antiferromagnetic intersite interaction between the Pr ions.
The effective magnetic moment of $\mu_{\rm eff}$ $=$ 3.61 $\mu_{\rm B}$/f.u. is close to 3.58 $\mu_{\rm B}$ for a free trivalent Pr ion.
On cooling below 4 K, $\chi(T)$ does not show divergent behavior but it is moderately suppressed, indicating that the CEF ground state is a van-Vleck paramagnetic one.

\begin{figure}
\centering
\includegraphics[width=17pc]{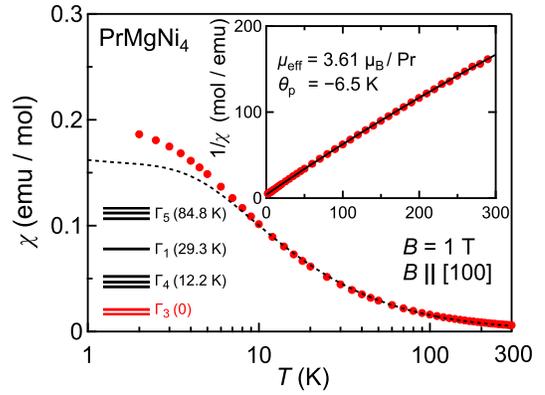}
\caption{(Color online) Temperature dependence of the magnetic susceptibility $\chi$ of PrMgNi$_4$ measured in the magnetic field of $B$ $=$ 1 T applied along the [100] direction. 
The dashed curve shows a calculation with considering the CEF levels and inter-site magnetic interaction of $K$ $=$ $-$0.3 K between the nearest Pr ions.
The inset shows $\chi^{-1}$, which was fitted with a modified Curie-Weiss equation. See text in detail.}
\label{f3}
\end{figure}

\begin{figure}
\centering
\includegraphics[width=16pc]{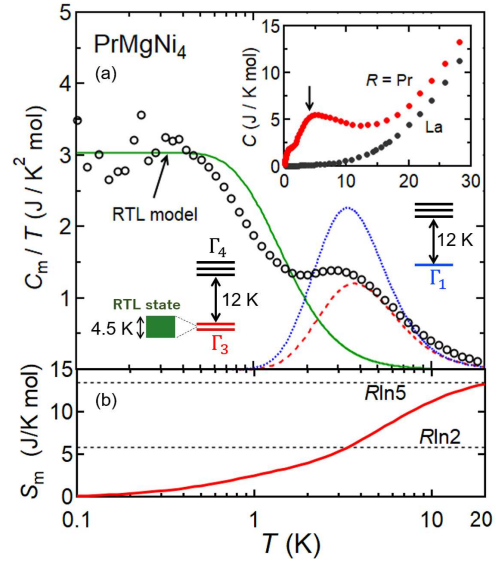}
\caption{(Color online) (a) Magnetic specific heat divided by temperature, $C_{\rm m}{/}T$, and (b) the magnetic entropy $S_{\rm m}$ of PrMgNi$_4$. The inset shows expanded data of $C(T)$ for $T$ $<$ 30 K. The arrow indicates the maximum at 4 K. The (red) dashed and (blue) dotted curves show calculations of the Schottky specific heat of the two-level systems for the doublet-triplet and singlet-triplet separated by12 K. The (green) solid curve presents the calculation with the model of the random two-level (RTL) assuming the energy width of 4.5 K.
}
\label{f4}
\end{figure}

The inset of Fig. \ref{f4} shows the temperature dependence of the specific heat $C$ of $R$MgNi$_4$ ($R$ $=$ Pr and La) for $T$ $<$ 30 K. 
For $R$ $=$ Pr,  a broad maximum appears at around 4 K, which results from the thermal excitation between the CEF levels as will be described later.
The electronic specific heat coefficient of $R$ $=$ La was estimated to be 19 mJ$/$(K$^2$  mol) by extrapolating the $C/T$ vs $T^2$ plot to $T$ $=$ 0 (not shown).

The magnetic specific heat divided by temperature, $C_{\rm m}{/}T$, of PrMgNi$_4$ is plotted against log$T$ for 0.1 $<$ $T$ $<$ 20 K in Fig. \ref{f4}(a).
The absolute value of $C_{\rm m}$ was estimated by subtracting the $C$ data of $R$ $=$ La as the lattice contribution from the measured $C$ data.
The broad maximum appearing at around 3 K can be moderately reproduced by the two-level model as follows.
The calculation with a doublet-triplet two-level model with an energy split of 12 K is shown with the (red) dashed curve ($\Gamma_{3}$), which agrees with the $C_{\rm m}{/}T$ data rather than that with a singlet-triplet model shown with the (blue) dotted curve ($\Gamma_{1}$). 
The respective two-level schemes are depicted in the inset.
This result indicates that the nonmagnetic CEF ground state is not the $\Gamma_{1}$ singlet but the $\Gamma_{3}$ doublet. 
On cooling below 3 K, the values of $C_\mathrm{m}$/$T$ are much larger than the calculated curve, suggesting the release of the magnetic entropy of the $\Gamma_{3}$ doublet at lower temperatures.

\begin{figure}
\centering
\includegraphics[width=18pc]{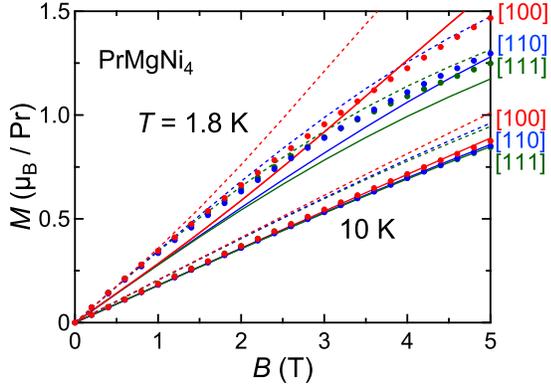}
\caption{(Color online) Isothermal magnetization $M(B)$ of PrMgNi$_4$ at 1.8 and 10 K in magnetic fields applied along the [100], [110], and [111] directions.
The dashed curves are calculations with the CEF parameters of $W$ $=$ $-$2.0 K and $x$ $=$ 0.77. The solid curves are those with the same parameters including intersite magnetic interaction of $K$ $=$ $-$0.3 K. 
}
\label{f5}
\end{figure}

The isothermal magnetization $M(B)$ data of PrMgNi$_4$ at $T$ $=$ 1.8 and 10 K in the magnetic fields along the [100], [110], and [111] directions are shown in Fig. \ref{f5}.
The $M(B)$ data at 1.8 K present anisotropic $B$-dependence; $M_{[100]}$ $>$ $M_{[110]}$ $>$ $M_{[111]}$.
This anisotropy of $M(B)$ can be explained by considering the Zeeman effect through the off-diagonal components between 
the $\Gamma_3$ doublet ground state and the first-excited $\Gamma_4$ triplet.  
In addition, the energy gap between them is about 12 K as indicated by the Schottky-type broad maximum of $C_{\rm m}/{T}$ at 4 K shown in Fig. \ref{f4}. 
Thereby, taking a diagram of the eigenstates for $J$ $=$ 4 under the cubic point group into account,\cite{Lea62}
CEF parameters of $W$ $=$ $-$2.0 K and $x$ $=$ 0.77 were proposed for the cubic CEF Hamiltonian, 
\begin{equation*}
\mathcal{H}_{\rm CEF}=W[\frac{x}{60}(O_{4}^{0}+5O_{4}^{4})+\frac{1-|x|}{1260}(O_{6}^{0}-21O_{6}^{4})].
\end{equation*}
As shown in Fig. \ref{f5}, the $M(B)$ data are smaller than the dashed curves calculated with the above CEF parameters.

To fit the data better, we take an intersite magnetic interaction into consideration using the following Hamiltonian.
\begin{equation}
\mathcal{H} = \mathcal{H}_{\rm CEF} + g_{J}\mu_{\rm B}\textbf{\textit J}\textbf{\textit B} - K\langle\textbf{\textit J}\rangle\textbf{\textit J},
\end{equation}
where $g_{J}=$  4$/$5 is the Land$\acute{\rm e}$ $g$-factor for a Pr$^{3+}$ ion, \textit{\textbf J} a total angular momentum, and $K$ a coefficient of the magnetic inter-site interaction between the Pr ions. 
The solid curves in Fig. \ref{f5} are the calculated $M(B)$ with a parameter of $K$ $=$ $-$0.3 K. 
The $M(B)$ data at 10 K are reproduced well, although those at 1.8 K show positive deviation from the calculated curve, e.g., for $B$ $<$ 3 T along the [100] direction, which is consistent with the enhanced $\chi(T)$ data observed for $T$ $<$ 10 K as shown in Fig. \ref{f3}.
In addition, the $M(B)$ data shows anisotropy at 1.8 K, however, it is much smaller than the calculation, which may result from the symmetry lowering of the Pr site due to the substituted Mg atoms as described above.
On heating to 10 K, the anisotropy in $M(B)$ almost fades out, probably because the splitting energy of the doublet could be overcome by the thermal excitations.
As shown in Fig. \ref{f3}, the $\chi(T)$ data for $T$ $>$ 10 K coincide with the dashed curve calculated by using the CEF parameters and the intersite magnetic interaction, whereas they deviate from the calculation for $T$ $<$ 10 K.
Similar deviation of the enhanced $\chi(T)$ from the calculation with a CEF scheme for $T$ $<$ 10 K was observed in PrMg$_3$, which was discussed with considering the possible $c$-$f$ hybridization effect.\cite{Morie09}


The magnetic entropy of the $\Gamma_3$ ground state doublet must be released on cooling by occurrence of a phase transition due to the degrees of freedom in the doublet. However, as shown in Fig. \ref{f4}(a), there appears no anomaly in $C_{\rm m}/T$ down to 0.1 K.
Instead, on cooling below 1 K, the $C_{\rm m}{/}T$ data increase and approach a constant value at $T$ $<$ 0.4 K, which is similar to those observed in the La substituted PrPb$_{3}$.\cite{Kawae01}
This behavior can be explained by a random two-level (RTL) model, where the density of states originating from the ground state doublet is uniformly distributed within a certain energy range.\cite{Anderson72,Phillips72}
The specific heat can be calculated by the RTL model with the following equation;
\begin{equation*}
	C_{\rm RTL}(T)= Nk_{\rm B} \int_0^\infty  n(E) (\frac{E}{k_{\rm B}T})^2 \frac{e^{-\frac{E}{k_{\rm B}T}}}{(1+e^{-\frac{E}{k_{\rm B}T}})^2} dE,
\end{equation*}
where the density of states of the ground state doublet $n(E)$ is assumed to be $1/\mathit{\Delta}$ in the energy range from zero to $\mathit{\Delta}$.
As shown with the (green) solid curve, the temperature variation of $C_{\rm m}{/}T$ that approaches the constant value at $T$ $<$ 0.4 K can be moderately reproduced by the RTL model with the energy split of 4.5 K, whose energy scheme is depicted in the inset.
The shoulder of $\rho(T)$ at $T$ $=$ 0.8 K shown with the (blue) arrow in the inset of Fig. \ref{f2} may be ascribed to the scattering of the conduction electrons by the thermal excitation between the distributed states. 
As shown in Fig. \ref{f4}(b), the magnetic entropy $S_{\rm m}$ was estimated by integrating the $C_{\rm m}{/}T$ data to the temperature.
The value of $S_{\rm m}$ reaches $R$ln2 at around 3 K, whose value is expected for the doublet ground state.

Taking the result of the single-crystal X-ray structural analysis into consideration, it is possible that the symmetry lowering of the Pr sites due to the Mg  atoms randomly substituted by 4.5\% for the Pr atoms splits the ground state doublet.
Thereby, the symmetry breaking of the cubic Pr site must hinder the long-range order of the quadrupolar degrees of freedom in the doublet.
It is noted that the AFQ orders in PrPb$_{3}$ and PrIr$_{2}$Zn$_{20}$ are collapsed by only a few percent of the La substitution.\cite{Kawae01,Mastumoto15}
Otherwise, strong hybridizations between the 4$f^2$ electron state and conduction bands play a role for the entropy release of the ground state doublet on cooling as was proposed in PrAg$_2$In and PrMg$_3$\cite{Yatskar96,Tanida06,Morie09}.
To explore the mechanism of the hindered quadrupole order and relevant quadrupole fluctuations, further study on single crystalline PrMgNi$_4$ samples with different amounts of the Mg substitution and the microscopic measurements such as neutron scattering, nuclear magnetic resonance, and muon spin relaxation are needed.


In summary, we have studied the structural, transport and magnetic properties of a cubic compound PrMgNi$_4$.
The single-crystal X-ray structural analysis revealed that the excess Mg atoms by 4.5\% are substituted for the Pr atoms at the 4$a$ site.
On cooling from 300 K, $\rho(T)$ monotonically decreases to 20 K and exhibits a shoulder at around 10 K. 
The $\chi(T)$ data follow the Curie-Weiss law of Pr$^{3+}$ ions.
A broad maximum in $C_{\rm m}/T$ at around 3 K can be reproduced by a two-level model with a ground state doublet and a first excited triplet.
On cooling below 1 K, $C_{\rm m}/T$ approach a constant value, which behavior is moderately explained by a random two-level model with an energy width of 4.5 K.
The twofold degeneracy of the non-magnetic ground state could be lifted by the symmetry lowering due to the excess Mg atoms substituted for the Pr atoms 
or strong hybridizations between the 4$f^2$ electron state and conduction bands, which prevents PrMgNi$_4$ from the long-range quadrupole ordering.



The authors would like to thank H. Funashima, H. Harima, H. S. Suzuki, K. Wakiya, R. J. Yamada, K. Urashima, K. T. Matsumoto and Y. Shimura for helpful discussion. 
The authors also thank Y. Shibata for the electron-probe microanalysis carried out at N-BARD, Hiroshima University. 
The measurements with MPMS, PPMS and the mFridge were performed at N-BARD, Hiroshima University. 
We acknowledge support from Center for Emergent Condensed-Matter Physics (ECMP), Hiroshima University. 
This work was financially supported by grants in aid from MEXT/JSPS of Japan, Grants Nos. JP26707017, JP15H05886 (J-Physics), and JP18H01182.




\end{document}